# Enhanced Control of High Harmonic Generation in Mixed Argon-Helium Gaseous Media

José Miguel Pablos-Marín[1,2,*], Javier Serrano [1,2, †], and Carlos Hernández-García[1,2]

[1] Grupo de Investigación en Aplicaciones del Láser y Fotónica, Departamento de Física Aplicada, Universidad de Salamanca, Pl. Merced s/n, E-37008 Salamanca, Spain

[2] Unidad de Excelencia en Luz y Materia Estructuradas, Universidad de Salamanca, E-37008, Salamanca, Spain

**Abstract** – High harmonic generation (HHG) in gaseous media provides a robust method for producing coherent extreme-ultraviolet (EUV) radiation and attosecond pulses. However, the spectral and temporal properties of these pulses—such as bandwidth and chirp—are fundamentally limited by the underlying generation mechanisms. Typically, tailoring the EUV emission involves modifying the properties of the intense infrared femtosecond driving pulse, and/or the macroscopic laser–matter configuration. Here, we focus on controlling the HHG process through the gas specie, introducing mixed-gas targets as a practical approach to enhance control over the EUV harmonic radiation. Through advanced simulations assisted by artificial intelligence that take into account both the quantum microscopic and macroscopic aspects of HHG, we demonstrate how mixtures of argon and helium modulate the emitted EUV harmonics. A simple model reveals that these modulations arise from coherent interference between harmonics emitted by different species at the single-atom level, and that they can be tuned by adjusting the macroscopic relative concentrations. Furthermore, by spatially separating the gas species into two distinct jets in a symmetric configuration, we gain additional control over the whole harmonic bandwidth. This strategy provides a realistic and versatile pathway to tailor EUV light and attosecond sources via HHG, while also enabling the identification of species-specific contributions to the process.

## 1. Introduction

The generation of coherent light in the extreme-ultraviolet (EUV) and soft X-ray regimes has seen significant progress over the past decades, primarily driven by the development of High Harmonic Generation (HHG) [1-4]. HHG is a highly-nonlinear, non-perturbative process that takes place when an intense femtosecond (fs) infrared (IR) laser pulse is focused in gaseous or solid targets. While recent advances have demonstrated HHG in bulk or thin crystal targets [5], gaseous media—such as gas jets, gas cells or gas-filled waveguides—remains as very efficient platforms for generating coherent EUV or soft x-ray radiation through HHG [6]. The underlying physics of HHG in gases can be intuitively described using a semiclassical model [7, 8]. Under the influence of the strong field of a fs IR laser pulse, an electronic wave-packet is first released through tunnel ionization. Secondly, the electronic wave-packet is accelerated by the laser field, and driven back towards the parent ion due to the oscillatory nature of the field. Finally, upon recollision, the kinetic energy gained by the electronic wave-packet during its excursion is emitted as high-frequency radiation. This three-step process repeats every half-cycle of the driving laser field, producing a train of attosecond pulses separated by half a cycle. As a consequence, the emitted radiation consists of a comb of odd-order harmonics of the fundamental frequency field with nearly equal intensity, leading to a *plateau* of harmonics that extend towards a sharp *cutoff* frequency. This contrasts with the rapidly decaying intensity characteristic of perturbative harmonic generation [9]. The

---

[*] Corresponding author: jmpablosm@usal.es

[†] Corresponding author: fjaviersr@usal.es

*cutoff* frequency, $\omega_{cutoff}$, is given by $\hbar\omega_{cutoff} = I_p + 3.17 U_p$, where $I_p$ is the atomic ionization potential of the atom —i.e. the energy necessary to extract the most energetic bound electron—, and $U_p$ is the ponderomotive energy, defined as $U_p = qE_0^2/4m\omega_0^2$, where $E_0^2$ and $\omega_0$ are the intensity and frequency of the driving laser pulse, respectively, and q and m the electron's charge and mass [10]. Thus, the maximum harmonic photon energy depends on the ionization potential [11], driving laser peak intensity [12], and driving laser wavelength [6].

The phase properties of the HHG spectrum are strongly influenced by the microscopic dynamics occurring at the atomic level. Typically, two main electronic trajectories contribute to the harmonic radiation within each half-cycle of the driving laser pulse, depending on the precise moment of ionization. These are commonly referred to as the short and long trajectories, according to the duration of the electron's excursion in the continuum. The short (long) trajectory contributes to a positive (negative) chirp into the harmonic emission. Thus, each trajectory imprints a characteristic intrinsic or dipole phase to the harmonic emission, which depends on the wave-packet excursion in the continuum before recombination [13]. This gives rise to the so-called *attochirp*, an intrinsic frequency chirp in the harmonic pulses associated to the trajectory-dependent phase variations [14-16]. Relevant to this work, the intrinsic dipole phase depends on the intensity of the driving laser pulse. The *attochirp* is regular enough to allow harmonic phase-locking, enabling the coherent synthesis of the higher-order harmonics into attosecond pulses [17, 18]. This distinctive property of the HHG emission has paved the way for applications time-resolved studies of electronic excitations [19], molecular photoionization dynamics [20-23], attosecond spectroscopy in solids [24], the observation of ultrafast magnetization dynamics [25], and x-ray diffractive imaging [26], among many others [4, 27, 28].

In an HHG experiment, the single-atom response is accompanied by macroscopic effects, since in a typical gas jet or gas cell, trillions of atoms contribute to the overall emission. As a result, the harmonic radiation from individual emitters coherently adds up, making phase-matching critical for determining the efficiency and characteristics of the generated harmonics [29-34]. For example, among the contributing electronic trajectories, short trajectories are more robust under phase-matching conditions and therefore tend to dominate the experimentally observed HHG radiation [29, 35]. In summary, both the microscopic (single-atom response) and macroscopic (phase-matching) properties must be considered to understand and control HHG.

Numerous strategies have been developed over the past decades to tailor the properties of the HHG radiation. For example, the cutoff energy can be extended through mixing different wavelengths [36], or by employing mid-IR driving lasers to access the soft x-ray regime [6, 37-39]. The spatial properties of the harmonics can be manipulated by sculpting the wavefront of the driving beam [40-43]. Phase-matching conditions can be engineered within the generation medium to enable the production of isolated attosecond pulses [44-46]. Furthermore, structuring the driving laser field makes it possible to imprint orbital angular momentum into the harmonics [47-50], or to generate circularly polarized harmonics [51-54].

In this work we explore the use of mixed gases—argon (Ar) and helium (He)—in HHG to enhance control over the spectral properties of the emitted harmonics. The use of gas mixtures offers an alternative way to tailor the HHG spectrum due to their distinct ionization potential, $I_p$. The $I_p$ influences the harmonic phase via the intrinsic dipole phase, meaning that the same harmonic can exhibit different phases when generated in different atomic species. This species-dependent phase variation has been experimentally measured through harmonic ellipsometry [55]. Previous studies have demonstrated the utility of mixed gases in HHG. For instance, a He-Ne mixture was employed to probe the harmonic phase and investigate the attosecond electronic dynamics underlying HHG [56]; the influence of Ar-Kr mixtures on the HHG spectrum was identified [57]; combinations of atomic and molecular gases have been used to retrieve molecular structures [58, 59], to generate near-circularly polarized attosecond pulses [60, 61], and more generally, to manipulate the harmonic emission [62]. Similar interference effect in the HHG spectrum due to the effect of subsequent ionized species have been observed in UV-driven HHG [11].

Here we propose a controlled mixture of Ar and He at specific concentrations to induce a tunable spectral gap in the HHG emission. Using advanced numerical simulations assisted by artificial intelligence (AI), which account for both the quantum microscopic (single-atom response) and macroscopic (phase-matching) aspects of HHG, we demonstrate how such gas mixtures modulate the emitted EUV HHG spectrum. A simple model reveals that these modulations result from coherent interference between harmonics generated by different species, and that their characteristics can be adjusted by varying the relative gas concentrations. Additionally, we demonstrate that spatial separation of the gas species in two jets introduces further control over the entire harmonic bandwidth through the Gouy phase of the driving laser beam. We thus provide a practical and versatile strategy for tailoring EUV and attosecond sources via HHG.

## 2. Material and Methods

The simulation setup of HHG in a mixed gas jet is illustrated in Fig. 1 a). An intense fs IR laser beam is focused into a gas jet composed of a mixture of Ar and He atoms. The mixture is characterized by the He concentration parameter $\eta_\%$, defined as

$$\eta_\% = \frac{n_{He}}{n_{He} + n_{Ar}} \times 100\% \quad (1)$$

where $n_{He}$ and $n_{Ar}$ are the densities of He and Ar gases, respectively. The Ar concentration is therefore $100 - \eta_\%$.

HHG is computed at every atom position within the jet, and the resulting harmonic field is propagated towards the far-field detector. The numerical simulations account for both the microscopic (single-atom response) and the macroscopic (phase-matching) physics of HHG. To model the microscopic response, we employ an AI-based method developed in our group [63]. Specifically, a neural network is trained to predict the dipole HHG acceleration at each atom position, using data generated from solving the three-dimensional time-dependent Schrödinger equation (3D-TDSE) under the single active electron approximation. The far-field macroscopic HHG emission that accounts for phase-matching effects is calculated using the integral solution of Maxwell's equations [64]. This hybrid AI-based macroscopic HHG method has been validated against both full numerical simulations [63], and experimental measurements of HHG driven by Hermite-Gauss beams [43].

In this work, two neural networks were independently trained and validated using datasets generated from 3D-TDSE simulations of HHG in Ar and He. The datasets comprised $4 \times 10^4$ data for each specie. In all cases, the driving pulse was modelled with a $\sin^2$ envelope, a central wavelength of 800 nm, and pulse duration of 7.7 fs in full width at half maximum in intensity. The datasets were generated covering realistic ranges of peak intensities and spatial phases. Specifically, the datasets spanned a range of spatial phases from 0 to $2\pi$, and peak intensities from $3.8 \times 10^{13}$ to $3.5 \times 10^{14}$ W/cm$^2$. The 3D-TDSE simulations were performed using a temporal grid of 8192 points with a resolution of 0.94 attoseconds for Ar and 16384 points with a resolution of 0.47 attoseconds for He. The spatial grids in cylindrical coordinates consisted of $2 \times 10^3$ points along the polarization axis and $8 \times 10^2$ points in the radial direction, with a resolution of $5.3 \times 10^{-3}$ nm for both atomic species. Under these conditions, the noise level in the normalized HHG spectra was approximately $10^{-4}$ relative to the plateau harmonics for Ar, and $10^{-5}$ for He. Training and validation of the neural networks were carried out using the Keras and TensorFlow libraries [65], employing the Adam optimizer and the mean squared error (MSE) as the loss function. The final trained models achieved an error metric (MSE) on the order of $10^{-6}$ when compared with direct 3D-TDSE outputs. Further technical details on the networks' architecture, training procedure, validation strategy and computational efficiency can be found in [63].

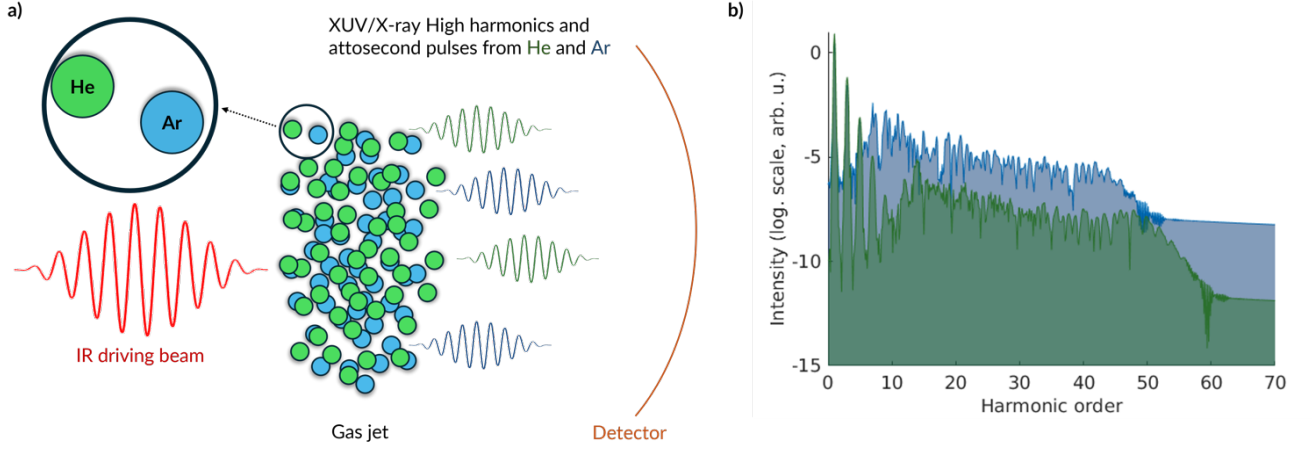

**Fig. 1. a)** Schematic of HHG in a mixed-gas jet composed of He and Ar. HHG occurs at each atom and the generated EUV radiation propagates to a far-field detector, where interference between the harmonic emissions from the different species becomes evident. **b)** Single-atom HHG spectra for He (green) and Ar (blue), obtained by solving the 3D-TDSE for a driving pulse with 800 nm wavelength, $2.84 \times 10^{14}$ W/cm$^2$ peak intensity, and 7.7 fs pulse duration.

In our HHG calculations we considered a Gaussian beam with beam waist of 30 μm, and peak intensity of $2.84 \times 10^{14}$ W/cm$^2$, focused into a low-density mixed gas jet (5 Torr total density). We have considered an infinitely thin gas jet placed at the axial position $z_0$. Figure 1(b) shows the computed single-atom HHG spectra obtained from 3D-TDSE simulations at the peak intensity of $2.84 \times 10^{14}$ W/cm$^2$ for He (green) and Ar (blue). Two features can be observed. First, the HHG yield from Ar is significantly stronger than from He. Second, the harmonic cutoff frequency is noticeably higher for He compared to Ar. These differences arise from the disparity in ionization potentials: He has a higher ionization potential ($I_{p,He}$=29.6 eV) than Ar ($I_{p,Ar}$=15.8 eV). On the one hand, Ar's lower ionization potential threshold leads to more efficient tunnel ionization, and hence, higher HHG yield. On the other hand, according to the three-step model, the harmonic cutoff frequency scales with the ionization potential, so He produces higher $\omega_{cutoff}$.

To gain physical insight beyond the advanced numerical HHG simulations, we also employed a simplified semiclassical Thin Slab Model (TSM) of HHG [66, 67], adapted to our mixed-gas configuration. The TSM describes the electric field of the $q$-th order harmonic, $E_q^j$ at position $(x, y, z_0)$, generated by atomic species $j \in \{Ar, He\}$, in terms of the amplitude, $|E_{fund}|$, and phase, $\phi_{fund}$, of the fundamental driving field as

$$E_q^j(x, y, z_0) = |E_{fund}(x, y, z_0)|^p e^{-iq\phi_{fund}(x,y,z_0) - i\phi_{dp}^j(x,y,z_0)} \quad (2)$$

where the harmonic amplitude scales as the $p$-power of the driving field, and $\phi_{dp}^j$ denotes the intrinsic dipole phase, which is element-specific. Relevant to our work, we assume the same $p$-power for both gases ($p$=4 [67]), whereas the intrinsic dipole phase is calculated for the short trajectory contributions, taking into account the distinct $I_p$ of Ar and He [31]. The TSM considers HHG emission from a thin slab placed at $z_0$. Thus, Eq. (2) serves as the source term, and the far-field harmonic emission is then computed via Fraunhofer propagation, enabling direct comparison with our advanced AI-based 3D-TDSE macroscopic numerical simulations. To model the mixed-gas geometry, the total harmonic field is calculated from the individual atom contributions as:

$$E_q(x, y, z_0) = \frac{\eta_\%}{100} E_q^{He}(x, y, z_0) + \left(1 - \frac{\eta_\%}{100}\right) E_q^{Ar}(x, y, z_0) \quad (3)$$

where $E_q^{He}$ and $E_q^{Ar}$ are the He and Ar contributions given by Eq. (2), differing only in their respective intrinsic dipole phase term.

## 3. Results

Figure 2a) presents the simulated far-field HHG spectra for different Ar/He gas mixtures, plotted as a function of He concentration $\eta_\%$, using the AI-based 3D-TDSE macroscopic HHG model. The mixed gas jet is placed at the focal plane, $z_0=0$. As anticipated from the single-atom HHG results shown in Fig. 1b), the overall harmonic yield is highest for pure Ar ($\eta_\% =0$), and lowest for pure He ($\eta_\%=100$). Interestingly, a regime of comparable HHG contributions from both gases emerges for $\eta_\%>80$, suggesting the possibility of interference effects.

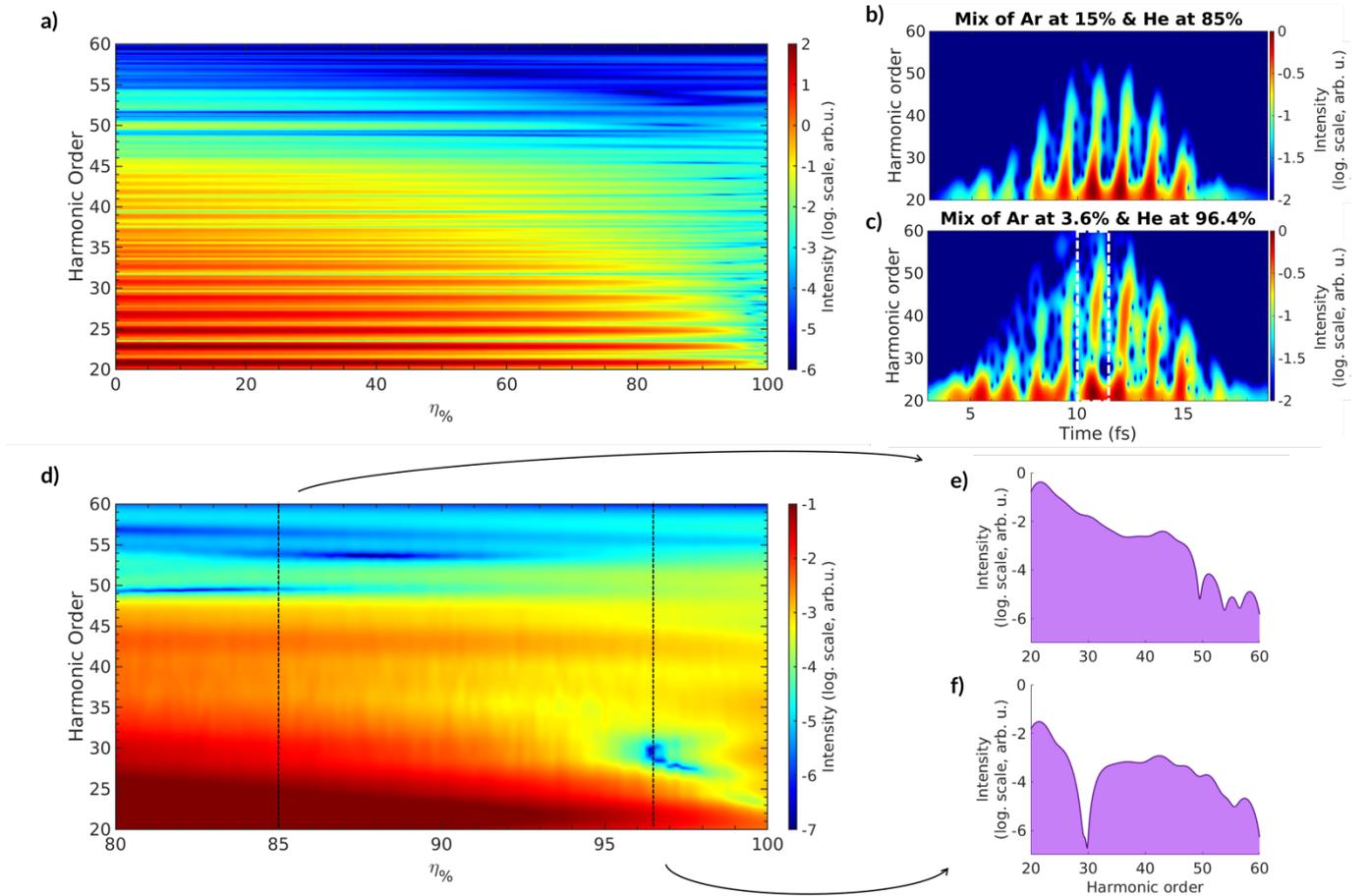

**Fig. 2. a)** HHG spectrum from macroscopic simulations as a function of helium percentage in the gas jet. The left end corresponds to pure argon (100% Ar), while the right end represents pure helium (100% He). Time-frequency analyses for two specific mixtures are shown in **b)** $\eta_\%=85$ (85% He / 15% Ar) and **c)** $\eta_\%=96.4$ (96.4% He / 3.6% Ar). In the latter, a modulation appears in the central attosecond burst (highlighted with a white dashed contour), which is further analyzed in the following panels. **d)** HHG spectrum filtered within the temporal window of the central burst shown in **c)**, applied across a reduced range in **a)**. Panels **e)** and **f)** display the filtered spectra for the two specific mixtures: **e)** $\eta_\%=85$ and **f)** $\eta_\%=96.4$. In **f)**, a pronounced minimum near the 30th harmonic order is clearly observed.

To gain further insight into the HHG emission dynamics, we performed a time-frequency analysis (also known as Gabor spectrogram), using a Gaussian spectral window of width $6\omega_0$. The time-frequency analyses for $\eta_\%=85$ and $\eta_\%=96.4$ are shown in Figs. 2b) and 2c), respectively. In both cases, the emission consists of a train of attosecond bursts with a positive temporal slope, indicative of short-trajectory contributions, and thus, positive *attochirp*. At $\eta_\%=85$ (Fig. 2b), the spectrogram exhibits smooth, regular emission features typical of standard HHG. In contrast, at $\eta_\%=96.4$ (Fig. 2c) a pronounced modulation appears. Specifically, there is a strong suppression of the HHG signal at the central burst, around the 30th harmonic order (white dashed line). This suppression points to destructive interference between the HHG contributions from He and Ar, that is frequency dependent on both harmonic order and driving intensity, varying from pulse to pulse.

To analyze this interference in more detail, we isolate the central attosecond burst by selecting the harmonic signal emitted between 10.4 fs and 11.2 fs. Experimentally, this could be achieved using few-cycle driving pulses or via spatial separation of the attosecond bursts using the attosecond lighthouse effect [68]. Alternatively, one could engineer the temporal envelope of the driving pulse as a trapezoidal function to equalize the dipole phase across the train. The resulting HHG spectra after selecting the central burst, and plotted as a function of $\eta_\%$ in the region of interest ($\eta_\%>80$) is shown in Fig. 2d). A notable suppression of the harmonic yield is observed around the 30th harmonic within the range $95<\eta_\%<98$. This is further illustrated in the HHG spectra shown in Figs. 2e) and 2f), corresponding to $\eta_\%=85$ and $\eta_\%=96.4$, respectively. These simulation results, obtained with the AI-based 3D-TDSE macroscopic model, reveal the possibility to tailor the HHG content through the mixture concentration.

To gain further insight into the results from advanced numerical simulations, and to better understand how to control the HHG spectrum via gas mixtures, we also performed simulations using the semiclassical TSM model. Figures 3a) and 3b) show the intensity of the 29th and 37th harmonics, respectively, as a function of the He concentration, $\eta_\%$, and the driving field peak intensity $|E_{fund}|^2$. The harmonic field is calculated using Eq. (3), noting that—unlike in the 3D-TDSE simulations—here the amplitudes of the HHG signal from Ar and He are assumed to be identical, so their contributions differ only in their intrinsic dipole phase.

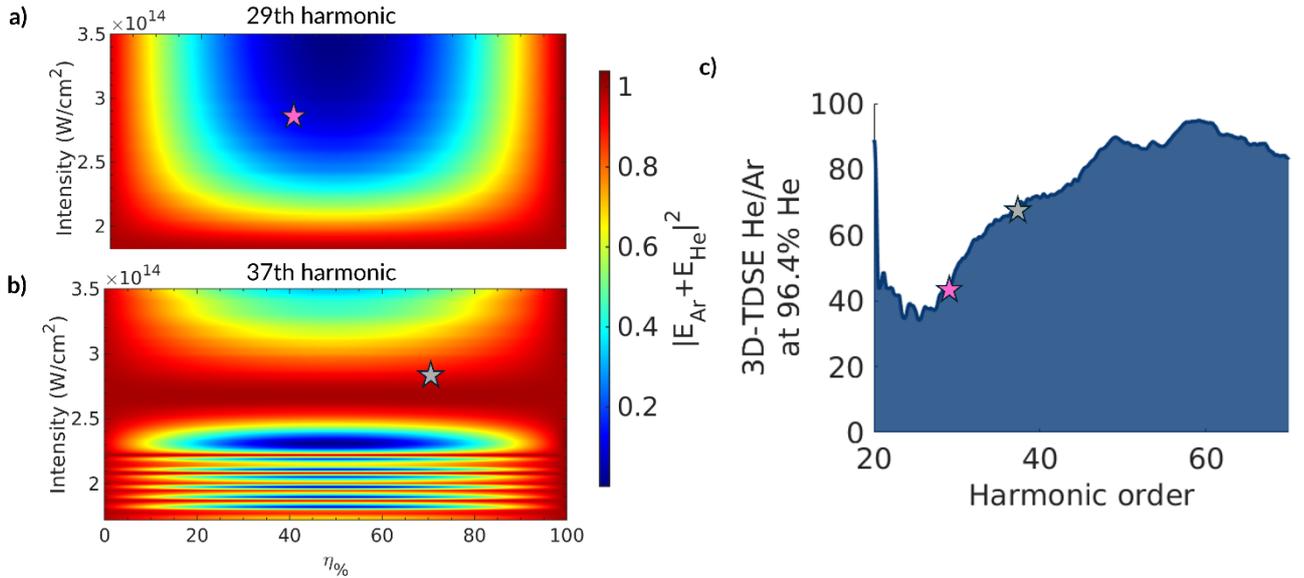

**Fig. 3. a) b)** HHG yield of the 29th and 37th harmonic, respectively, obtained from the TSM approach—Eq. (3)—, as a function of driving peak intensity and He concentration $\eta_\%$. **c)** Ratio between the single-atom He and Ar HHG spectrum obtained within the 3D-TDSE. The pink and grey stars denote the ratio obtained for the 29th and 37th harmonic orders, to identify their specific contributions in panels **a)** and **b)**.

Figures 3a) and 3b) illustrate how the total harmonic yield is strongly modulated due to the distinct intrinsic dipole phase from each contribution. This modulation is highly sensitive to both the harmonic order and the driving field peak intensity. This explains the features observed in Fig. 2c), where the suppression of the harmonic yield varies with the harmonic order, and differs from burst to burst, since the local driving field intensity changes within the pulse.

To clarify the specific case observed in Figs. 2c) and 2f) for $\eta_\%$=96.4, Fig. 3c) shows the ratio of the 3D-TDSE single-atom HHG yield from He and Ar as a function of the harmonic order. Specifically, for the 29$^{th}$ harmonic (pink star), the ratio is ~40%, while for the 37$^{th}$ harmonic (grey star), ~70%. Considering that the peak intensity of such particular case is $2.84 \times 10^{14}$ W/cm$^2$, we can locate this specific condition in Figs. 3a) and 3b), indicated by the corresponding pink and grey stars. This comparison clearly shows that under these conditions, the 29$^{th}$ harmonic yield is strongly suppressed due to the destructive interference between the He and Ar contributions, whereas the 37$^{th}$ harmonic is barely affected. This features are well reproduced by the advanced numerical simulations (Fig. 2f), demonstrating that the strong modulation in the HHG spectrum arises from the interference between the two species. This interference is directly linked to the different ionization potentials of the two species, which imprint a distinct intrinsic dipole phase.

Building on the previous results, and in order to extend the degree of control over HHG emission using gas mixtures, we propose introducing the Gouy phase as an additional tuning parameter. A Gaussian beam experiences a phase shift, known as Gouy phase, as it propagates through the focus within the Rayleigh range. As such, by spatially separating the Ar and He gases into two distinct jets that are symmetrically displaced from the focal plane, an extra phase term is introduced, enabling further control of the interference between the two HHG contributions. This symmetric configuration ensures that the driving field peak intensity remains the same in the two jets. Thus, relative to the previous results, the main difference arises from the Gouy phase shift. This arrangement is depicted in Fig. 4a), where the parameter $\Delta z$ denotes the symmetric axial displacement of each gas jet center relative to the focal plane.

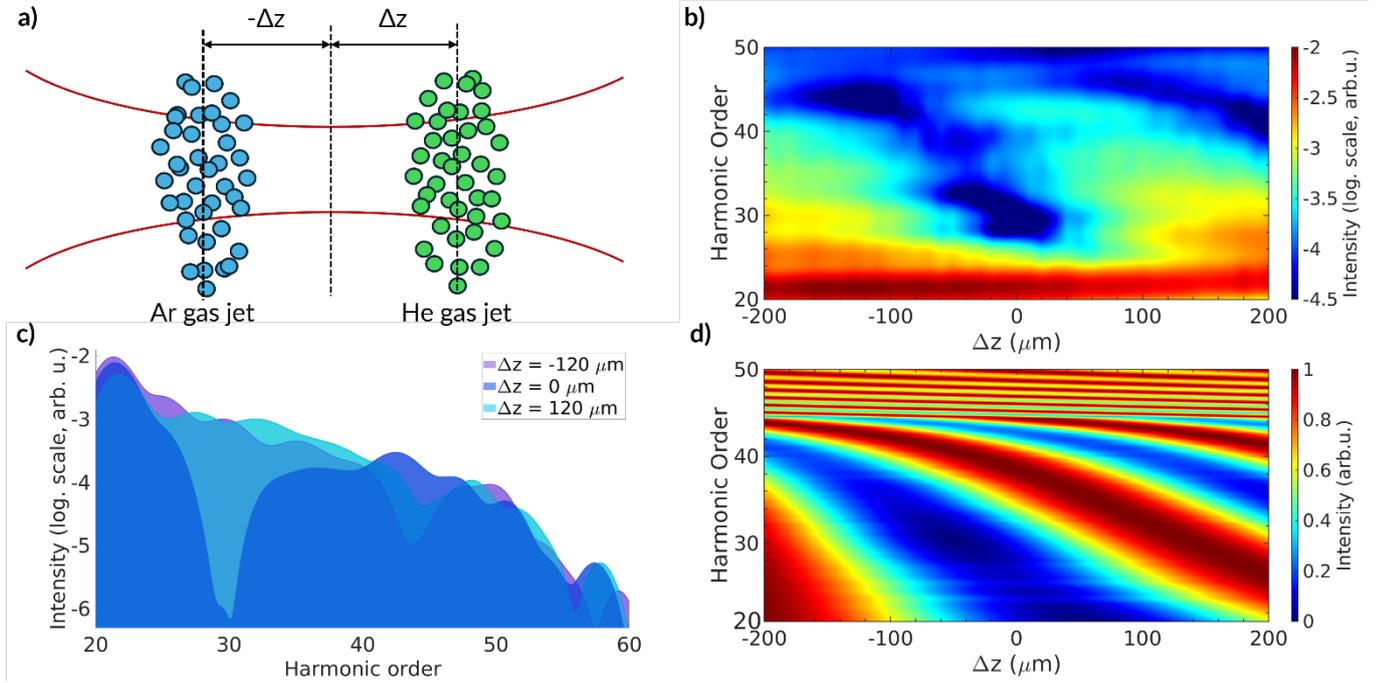

**Fig. 4. a)** Schematic of HHG in a mixed-gas jet composed of two separate Ar and He jets that are symmetrically displaced by $\Delta z$ from the driving field focal plane. **b)** HHG spectral intensity from macroscopic simulations as a function $\Delta z$. **c)** HHG spectra for three specific cases at $\Delta z$=-120 μm, 0 μm, and 120 μm. **d)** HHG spectral intensity from the TSM, for a driving peak intensity of $2.84 \times 10^{14}$ W/cm$^2$, reproducing the numerical advanced result simulations presented in **b)**.

Figure 4b) shows the results of the AI-assisted 3D-TDSE macroscopic simulations of the far-field HHG emission as a function of the gas jet displacement $\Delta z$, for a fixed He concentration of $\eta_\%$=96.4. At $\Delta z$=0, the results match those presented in Fig. 2f). Note that each gas jet is modelled as an infinitely thin layer located at a specific z-position, and, as in Fig. 2d), the analysis isolated the central attosecond burst of the HHG train. As seen in Fig. 4b), the interference pattern can be actively tuned by the gas jet displacement, enabling the suppression or enhancement to span over the entire HHG spectrum. As an illustration, Fig.

4c) shows three example harmonic spectra corresponding to Δz=-120 µm, 0 µm, and 120 µm. The gas jet displacement allows to tune the harmonic spectrum over the entire HHG bandwidth.

Finally, Fig. 4d) shows the TSM model results for a driving peak intensity of $2.84 \times 10^{14}$ W/cm$^2$, which reproduce the trend observed in the advanced numerical simulations. This agreement confirms the capability of the simplified TSM approach to provide clear physical insight into how gas mixtures can be engineered to tailor the spectral content of high-order harmonic emission.

## 4. Discussion

We have demonstrated the role that mixing atomic species plays in HHG. In particular, we have shown how the interference between the high-order harmonic radiation emitted between different species can be exploited to tailor the HHG spectrum. Although this interference originates at the microscopic single-atom HHG response, we have demonstrated through advanced numerical simulations that it survives under macroscopic phase-matching conditions in low-density gas jets. We have shown the effect considering Ar and He mixtures, though our work could be extended to other noble gas species, where their different ionization potential would play the relevant role for the observed interference effect.

We have demonstrated that this interference effect can be observed by selecting one burst from the attosecond pulse train. However, strategies as the implementation of attosecond lighthouse effect [68], or the use of few-cycle driving pulses provide alternative practical routes to observe this effect without temporal filtering. Moreover, shaping the driving laser pulse into a trapezoidal temporal envelope could further enhance the visibility of the interference effect, remaining similar across all the pulses within the attosecond pulse train.

We have also shown that this interference effect can be effectively controlled by spatially separating the gas mixture into two homogeneous gas jets symmetrically displaced with respect to the focal plane. This contribution introduces a tunable modulation across the entire HHG spectrum, providing a flexible method for tailoring and shaping the spectral content of the HHG emission. Currently, spectral selection of the harmonics is primarily achieved through thin metallic filers, typically Al, Ti, Zr, Ni or Ag. While straightforward to implement, such filters offer limited flexibility for spectral shaping of the attosecond HHG emission. Yet, the shortest attosecond pulses reported to date have employed metallic filters as 400-nm tin [69], 100-nm Zr [70], or 200-nm Al, Zr and Ag [46]. In this context, we provide the use of two separated gas jets of different homogeneous as a promising route to precisely tailor the EUV bandwidth of the HHG attosecond emission.

In addition, our proposal may provide a new degree of freedom for controlling the generation of circularly polarized HHG attosecond radiation from two sources [71, 72]. Finally, although our work focuses on the use of low density gas jets, it is well known that gas cells or waveguides can enhance the HHG yield through proper phase-matching conditions [73]. In such a case, the phase-matching conditions would vary for each atomic element, and this must be carefully considered. Nevertheless, our findings pave the route to explore combinations of longer and denser media with structured fields and tailored gas mixtures as a promising pathway for enhanced and tunable EUV attosecond emission.

**Funding**

This project has received funding from the European Research Council (ERC) under the European Union's Horizon 2020 research and innovation program (Grant Agreement No. 851201), and from the Department of Education of the Junta de Castilla y León and FEDER Funds (Escalera de Excelencia CLU-2023-1-02 and grant No. SA108P24). We acknowledge support from Ministerio de Ciencia e Innovación (PID2022-142340NB-I00). The authors thankfully acknowledge RES resources provided by BSC in MareNostrum 5, and CESGA in Finisterrae 3 to FI-2024-3-0035.

**Conflicts of interest**

The authors have nothing to disclose.

**Data availability statement**

Data underlying the results presented in this paper may be obtained from the authors upon reasonable request.

**Author contribution statement**

Conceptualization, all authors; Methodology, all authors; Software, J.M.P.-M., J. S.; Validation, J.M.P.-M., J. S.; Formal Analysis, J.M.P.-M.; Investigation, all authors; Resources, C.H.-G.; Data Curation, J.M.P.-M; Writing – Original Draft Preparation, all authors; Writing – Review & Editing, all authors; Visualization, all authors; Supervision, J. S. C.H.-G.; Project Administration, C.H.-G.; Funding Acquisition, C.H.-G.

J.M. Pablos-Marín et al.: Enhanced Control of High Harmonic Generation in Mixed Argon-Helium Gaseous Media[20] Calegari F, Trabattoni A, Palacios A, Ayuso D, Castrovilli MC, Greenwood JB, Decleva P, Martín F, Nisoli M, Charge migration induced by attosecond pulses in bio-relevant molecules, J. Phys. B: At. Mol. Opt. Phys. 49 142001 (2016). https://doi.org/10.1088/0953-4075/49/14/142001

[21] Beaulieu S, Comby A, Clergerie A, Caillat J, Descamps D, Dudovich N, Fabre B, Géneaux R, Légaré F, Petit S, Pons B, Porat G, Ruchon T, Taïeb R, Blanchet V, Mairesse Y, Attosecond-resolved photoionization of chiral molecules, Science 358 (6368) (2017) 1288–1294. https://www.science.org/doi/abs/10.1126/science.aao5624

[22] Grundmann S, Trabert D, Fehre K, Strenger N, Pier A, Kaiser L, Kircher M, Weller M, Eckart S, Schmidt LPH, Trinter F, Jahnke T, Schöffler MS, Dörner R, Zeptosecond birth time delay in molecular photoionization, Science 370 339–341 (2020). https://doi.org/10.1126/science.abb9318

[23] Borrego-Varillas R, Lucchini M, Nisoli M, Attosecond spectroscopy for the investigation of ultrafast dynamics in atomic, molecular and solid-state physics, Rep. Prog. Phys. **85** 066401 (2022) https://doi.org/10.1088/1361-6633/ac5e7f

[24] Tao Z, Chen C, Szilvási T, Keller M, Mavrikakis M, Kapteyn H, Murnane M, Direct time-domain observation of attosecond final-state lifetimes in photoemission from solids, Science 353 62–67 (2016). https://doi.org/10.1126/science.aaf6793

[25] Tengdin P, You W, Chen C, Shi X, Zusin D, Zhang Y, Gentry C, Blonsky A, Keller M, Oppeneer PM, Kapteyn HC, Tao Z, Murnane MM, Critical behavior within 20 fs drives the out-of-equilibrium laser-induced magnetic phase transition in nickel, Sci. Adv. 4 (3) (2018) eaap9744. https://doi.org/10.1126/sciadv.aap9744

[26] Miao J, Ishikawa T, Robinson IK, Murnane MM, Beyond crystallography: Diffractive imaging using coherent x-ray light sources, Science 348 (6234) (2015) 530–535. https://doi.org/10.1126/science.aaa1394

[27] Shi X, Liao CT, Tao Z, Cating-Subramanian E, Murnane MM, Hernández-García C, Kapteyn HC, Attosecond light science and its application for probing quantum materials, J. Phys. B: At. Mol. Opt. Phys. 53 (18) (2020) 184008. https://doi.org/10.1088/1361-6455/aba2fb

[28] Midorikawa K, Progress on table-top isolated attosecond light sources, Nat. Photon. 16 (4) (2022) 267–278. https://doi.org/10.1038/s41566-022-00961-9

[29] Salieres P, L'Huillier A, Lewenstein M, Coherence Control of High-Order Harmonics, Phys. Rev. Lett. 74, 3776. (1995). https://doi.org/10.1103/PhysRevLett.74.3776

[30] Rundquist A, Durfee CG, Chang Z, Herne C, Backus S, Murnane MM, Kapteyn HC. Phase-Matched Generation of Coherent Soft X-rays. Science **280**,1412-1415(1998). https://doi.org/10.1126/science.280.5368.1412

[31] Gaarde MB, Tate JL, Schafer KJ. Macroscopic aspects of attosecond pulse generation. J. Phys. B: At. Mol. Opt. Phys. 41 132001 (2008). https://doi.org/10.1088/0953-4075/41/13/132001

[32] Popmintchev T, Chen MC, Arpin P, Murnane MM, Kapteyn HC, The attosecond nonlinear optics of bright coherent X-ray generation, Nat. Photon. 4, 822 (2010) https://doi.org/10.1038/nphoton.2010.256

[33] Fu Z, Chen Y, Peng S, Zhu B, Li B, Martín-Hernández R, Fan G, Wang Y, Hernández-García C, Jin C, Murname MM, Kapteyn HC, Tao Z, Extension of the bright high-harmonic photon energy range via nonadiabatic critical phase matching. Sci. Adv.**8**,eadd7482(2022). https://doi.org/10.1126/sciadv.add7482

[34] Weissenbilder R, Carlström S, Rego L, Guo C, Heyl CM, Smorenburg P, Constant E, Arnold CL, L'huillier A, How to optimize high-order harmonic generation in gases. Nat Rev Phys 4, 713–722 (2022). https://doi.org/10.1038/s42254-022-00522-7

[35] Balcou P, Salieres P, L'Huillier A, Lewenstein M. Generalized phase-matching conditions for high harmonics: The role of field-gradient forces. Phys. Rev. A 55(4), 3204. (1997). https://doi.org/10.1103/PhysRevA.55.3204

[36] Chipperfield LE, Robinson JS, Tisch JWG, Marangos JP. Ideal waveform to generate the maximum possible electron recollision energy for any given oscillation period. Phys Rev Lett. **102**, 063003 (2009). https://doi.org/10.1103/PhysRevLett.102.063003

[37] Johnson AS, Austin DR, Wood DA, Brahms C, Gregory A, Holzner KB, Jarosch S, Larsen EW, Parker S, Strber CS, Ye P, Tisch JWG, Marangos J, High-flux soft x-ray harmonic generation from ionization-shaped few-cycle laser pulses. Sci. Adv.4,eaar3761(2018). https://doi.org/10.1126/sciadv.aar3761

[38] Takahashi EJ, Kanai T, Ishikawa KL, Nabekawa Y, Midorikawa K. Coherent water window x ray by phasematched high-order harmonic generation in neutral media. Phys Rev Lett. **101**, 253901 (2008). https://doi.org/10.1103/PhysRevLett.101.253901

[39] Cousin SL, Silva F, Teichmann S, Hemmer M, Buades B, Biegert J. High-flux table-top soft x-ray source driven by sub-2-cycle, CEP stable, 1.85-μm 1-kHz pulses for carbon K-edge spectroscopy, Opt. Lett. 39 (18), 5383-5386 (2014). https://doi.org/10.1364/OL.39.005383
11